\documentclass[twocolumn,nofootinbib,floatfix,aps,prd,10pt]{revtex4-2}
\usepackage[utf8]{inputenc}
\usepackage{graphics,graphicx}
\usepackage{amsmath, amssymb}
\usepackage{multirow}
\usepackage{color}
\usepackage{verbatim}
\usepackage{hyperref}
\usepackage[normalem]{ulem}  
\usepackage{ulem}
\usepackage{color}
\usepackage{cancel}
\usepackage{mathtools}
\usepackage{epstopdf}

\renewcommand\sout{\bgroup\color{blue} \ULdepth=-.5ex \ULset}
\def\slashchar#1{\setbox0=\hbox{$#1$}  
\dimen0=\wd0     
\setbox1=\hbox{/} \dimen1=\wd1  
\ifdim\dimen0>\dimen1   
\rlap{\hbox to \dimen0{\hfil/\hfil}} 
#1     
\else     
\rlap{\hbox to \dimen1{\hfil$#1$\hfil}} 
/      
\fi}

\newcommand{\dd}{\mathrm{d}}
\newcommand{\eps}{\epsilon}
\newcommand{\pp}{\partial}

\begin{document}
\title{Anatomy of critical fluctuations in hadronic matter}

\date{\today}
\author{Micha\l{} Marczenko}
\email{michal.marczenko@uwr.edu.pl}
\affiliation{Incubator of Scientific Excellence - Centre for Simulations of Superdense Fluids, University of Wroc\l{}aw, plac Maksa Borna 9, PL-50204 Wroc\l{}aw, Poland}

\begin{abstract}
Critical phenomena in phase transitions of strongly interacting matter, governed by quantum chromodynamics, are inherently encoded in the fluctuations of conserved charges. In this work, we study the net-baryon number density fluctuations, including the lowest-lying nucleon and the baryonic resonance $\Delta(1232)$, based on the parity doublet model in the mean-field approximation. We focus on the qualitative features of the second-order susceptibility of the net-baryon number density in dense hadronic matter and how the inclusion of $\Delta(1232)$ affects it. We demonstrate that the fluctuations of the individual baryons do not necessarily reflect the total net-baryon number fluctuations at finite density, due to the non-trivial correlations between different particle species. Our results highlight the role of baryonic correlations in the interpretation of data from heavy ion collision experiments.
\end{abstract}
\maketitle

\section{Introduction}
\label{sec:intro}

The exploration of critical phenomena in phase transitions of strongly interacting matter governed by quantum chromodynamics (QCD) is one of the goals of the current ultra-relativistic heavy ion collision (HIC) experiments. In particular, the central attention of the experiments at BNL and CERN and the upcoming large-scale nuclear experiments FAIR in GSI and NICA in Dubna is focused on understanding the properties of dense baryonic matter (see, e.g.,~\cite{Sorensen:2023zkk} for a recent review). Low-energy hadronic matter is expected to undergo a phase transition to quark-gluon plasma with increasing baryon density and/or temperature. From {\it{ab initio}} lattice QCD (LQCD) calculations at vanishing net-baryon density, we know that strongly interacting matter undergoes a simultaneous smooth restoration of chiral symmetry and deconfinement of hadronic matter to quark-gluon plasma at $T_c\simeq 0.155~$GeV~\cite{Bazavov:2014pvz, Borsanyi:2018grb, Bazavov:2017dus, Bazavov:2020bjn, Bazavov:2020bjn, Aoki:2006we}. 

The determination of the properties of the chiral phase transition in LQCD calculations at high net-baryon densities is currently unattainable due to a well-known sign problem. Nevertheless, many effective models predict a first-order transition at low temperatures (see, e.g.,~\cite{Bowman:2008kc, Ferroni:2010ct, Klevansky:1992qe, Buballa:2003qv}). This would imply the existence of a critical point on the QCD phase diagram. Despite experimental efforts at the Relativistic Heavy Ion Collider at BNL~\cite{STAR2010} and the Super Proton Synchrotron at CERN~\cite{Mackowiak-Pawlowska:2020glz}, its existence has so far not been confirmed~\cite{Bzdak:2019pkr}.

Fluctuations of conserved charges are excellent probes of critical behavior associated with phase transitions. The expectation is that they would grow rapidly and exhibit nonmonotonic behavior at the freeze-out in HICs~\cite{Bazavov:2012vg, Borsanyi:2014ewa, Karsch:2010ck, Braun-Munzinger:2014lba, Vovchenko:2020tsr, Braun-Munzinger:2020jbk} and the QCD phase boundary~\cite{Stephanov:1999zu, Asakawa:2000wh, Hatta:2003wn, Friman:2011pf}. Because of the statistical nature of HIC experiments, the fluctuations of conserved charges are connected to the event-by-event fluctuations. First indications of nonmonotonic behavior were seen in the net-proton multiplicity fluctuations in central Au+Au collisions in the BES-I program, which covered $\sqrt{s_{\rm NN}}=7.7-200~$GeV. Nevertheless, more precise data at low collision energies is needed for better statistics and determination of the behavior of higher-order multiplicity fluctuations.

Chiral symmetry is a fundamental symmetry of quantum QCD. One of the consequences of the dynamical restoration of chiral symmetry is the appearance of parity doubling of baryons, which has been verified in LQCD calculations at finite temperature and the vanishing baryon chemical potential for the octet and decouplet of light baryons~\cite{Aarts:2015mma, Aarts:2017rrl, Aarts:2018glk}. The masses of positive-parity baryons were found to be almost independent of temperature. At the same time, the masses of their negative-parity baryons were found to drop rapidly as the chiral symmetry was being restored, leaving the opposite-parity baryons degenerate with a finite mass in the vicinity of the chiral crossover. This phenomenon can be described in the effective framework of the parity doublet model, which is a generalization of the famous linear sigma model~\cite{Walecka:1974qa, Serot:1984ey}. The model has been applied to the phenomenology of HICs and studies of neutron stars~\cite{Detar:1988kn, Jido:1999hd, Jido:2001nt, Dexheimer:2007tn, Gallas:2009qp, Paeng:2011hy, Sasaki:2011ff, Gallas:2011qp, Zschiesche:2006zj, Benic:2015pia, Marczenko:2017huu, Marczenko:2018jui, Marczenko:2019trv, Marczenko:2020wlc, Marczenko:2020jma, Marczenko:2021uaj, Marczenko:2022hyt, Mukherjee:2017jzi, Mukherjee:2016nhb, Dexheimer:2012eu, Weyrich:2015hha, Sasaki:2010bp, Yamazaki:2018stk, Yamazaki:2019tuo, Ishikawa:2018yey, Giacosa:2011qd, Motohiro:2015taa, Minamikawa:2020jfj, Kong:2023nue, Fraga:2023wtd, Eser:2023oii, Gao:2024mew, Kong:2023nue, Minamikawa:2023ypn, Gao:2024chh}.

The effects of the restoration of chiral symmetry on the fluctuations of conserved charges have recently been studied in the context of the parity doublet model with nucleonic degrees of freedom~\cite{Marczenko:2023ohi, Koch:2023oez}. The differences in the qualitative critical behavior of opposite parity chiral partners and their correlations were shown to be non-trivial. In this work, we use the parity doublet model to calculate the susceptibilities of the net-baryon number distribution. We generalize the analysis done in~\cite{Koch:2023oez}, based on a framework that allows the evaluation of fluctuations in a single baryonic parity doublet, that is, the nucleon and its chiral partner. We explicitly include the $\Delta(1232)$ resonance and its chiral partner $\Delta(1700)$ within the framework of the extended parity doublet model~\cite{Takeda:2017mrm}. The inclusion of $\Delta$ matter introduces additional correlations that are expected to affect the fluctuations of conserved charges. We analyze the susceptibilities in individual chiral doublets and various correlations among them. We emphasize the low-temperature and high-density part of the model phase diagram. It should be noted that the consequences of the appearance of the $\Delta(1232)$ resonance are found to be crucial for the properties of the low-temperature equation of state (EOS) and the structure of neutron stars (see, e.g.,~\cite{Drago:2013fsa, Drago:2014oja, Li:2019tjx, Motta:2019ywl, Cai:2015hya, Ribes:2019kno, Marczenko:2021uaj, Marczenko:2022hyt}). Therefore, it is important to explore further the underlying role of the $\Delta(1232)$ resonance in dense matter under extreme conditions.

This work is organized as follows. In Sec.~\ref{sec:pd_model}, we introduce the hadronic parity doublet model for the nucleon and $\Delta(1232)$. In Sec.~\ref{sec:fluct}, we introduce the cumulants and susceptibilities of the net-baryon number. In Sec.~\ref{sec:results}, we present our results. Finally, Sec.~\ref{sec:results} is devoted to the conclusions.

\section{Parity doublet model}
\label{sec:pd_model}

In this section, we briefly review the $\rm SU(2)$ parity doublet for the nucleon and  $\Delta(1232)$ resonance~\cite{Takeda:2017mrm}. Here, we follow~\cite{Marczenko:2022hyt} assuming isospin symmetry. The thermodynamic potential of the model in the mean-field approximation reads
\begin{equation}
    \Omega = \sum_\alpha \Omega_{\alpha} + V_\sigma + V_\omega \rm,
\end{equation}
where $\alpha$ denotes spin-$1/2$ nucleons ($N_\pm$) and  spin-$3/2$ $\Delta$ resonances ($\Delta_\pm$), with $\pm$ denoting positive (negative) parity states. The mean-field potentials are
\begin{align}
    V_\sigma &= -\frac{\lambda_2}{2}\sigma^2 + \frac{\lambda_4}{4}\sigma^4 - \frac{\lambda_6}{6}\sigma^6 - \eps \sigma \rm,\\
    V_\omega &= -\frac{m_\omega^2}{2}\omega^2 \rm,
\end{align}
where $\lambda_2 = \lambda_4 f_\pi^2 - \lambda_6 f_\pi^4 - m_\pi^2$ and $\eps = m_\pi^2 f_\pi$. The masses of the $\pi$ and $\omega$ mesons are denoted as $m_\pi$ and $m_\omega$, respectively. The decay constant of the pion is denoted as $f_\pi$. The parameters $\lambda_4$, $\lambda_6$ are fixed to the properties of the nuclear ground state at vanishing temperature: the saturation density $n_0$, binding energy $E/A - m_{N_+}$, and the incompressibility $K$ (see Table~\ref{tab:gs}). An additional constraint on the model is imposed by the condition that $\Delta$ matter is not present in the nuclear ground state and subsaturation densities. Therefore, determining the ground state properties in the current model is done in the same way as in the pure nucleonic parity doublet model (see, e.g.,~\cite{Koch:2023oez}).

\begin{figure}
    \centering
    \includegraphics[width=\linewidth]{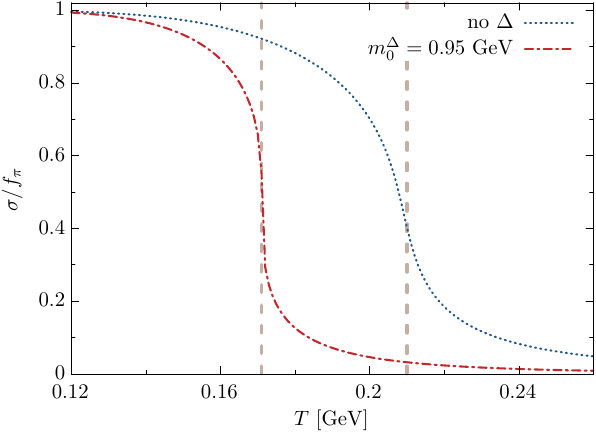}
    \caption{Normalized expectation value of the $\sigma$ mean field at vanishing baryon chemical potential for $m_0^N=0.8~\rm GeV$, and $m_0^\Delta=\infty$ (blue, dotted line) and $m_0^\Delta = 0.95$ (red, dash-dotted line). The vertical dashed lines mark the pseudocritical temperatures $T_c$.}
    \label{fig:sigma_T}
\end{figure}

\begin{figure}
    \centering
    \includegraphics[width=\linewidth]{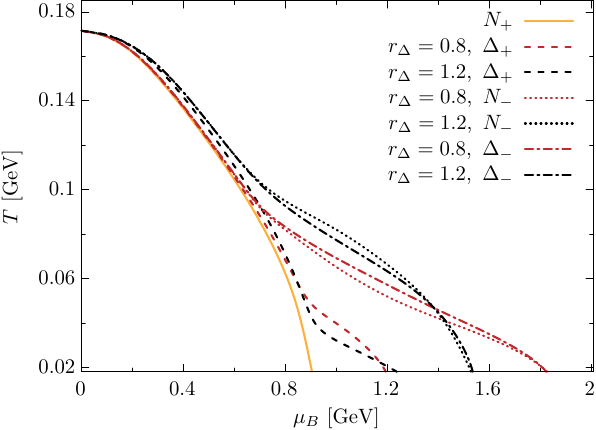}
    \caption{Phase diagram obtained in the parity doublet model. The lines are obtained from the minima of $\pp \sigma / \pp \mu_\alpha$ (see text for details). Note that the lines for $r_\Delta=1.2$ and $N_+$ practically overlap with the corresponding line for $r_\Delta=0.8$ and are not shown in the figure.}
    \label{fig:phase_diagram}
\end{figure}

The kinetic part of the thermodynamic potential reads
\begin{equation}\label{eq:kinetic_thermo}
    \Omega_\alpha = \gamma_\alpha T \int \frac{\dd^3p}{(2\pi)^3} \; \ln(1-f_\alpha) + \ln(1-\bar f_\alpha) \rm,
\end{equation}
where 
\begin{align}
         f_\alpha &= \left(1 + e^{(E_\alpha - \mu_\alpha)/T}\right)^{-1}\rm,\\
    \bar f_\alpha &= \left(1 + e^{(E_\alpha + \mu_\alpha)/T}\right)^{-1}\rm,
\end{align}
are the particle and antiparticle Fermi-Dirac distribution functions, respectively. The spin-isospin degeneracy factors are $\gamma_{N_\pm} = 2 \times 2 = 4$ and $\gamma_{\Delta_\pm} = 4 \times 4 = 16$. The dispersion relation is $E_\alpha = \sqrt{\boldsymbol p^2 + m_\alpha^2}$, the effective chemical potential $\mu_\alpha = \mu_B - g_\omega^\alpha \omega$, and $T$ is the temperature. We note that in the current model, the positive and negative parity states are coupled to the vector meson with the same strength, that is, $g_\omega^N = g_\omega^{N_+} = g_\omega^{N_-} $ and $g_\omega^\Delta = g_\omega^{\Delta_+} = g_\omega^{\Delta_-}$. Thus, we also define the effective chemical potentials $\mu_N = \mu_{N_+} = \mu_{N_-}$ and $\mu_\Delta = \mu_{\Delta_+} = \mu_{\Delta_-}$. In the literature, it is customary to parametrize the coupling of the $\omega$ meson to the $\Delta$ baryon in terms of the coupling to the nucleon:
\begin{equation}
   r_\Delta   = \frac{g_\omega^\Delta}{g_\omega^N}\rm.
\end{equation}
The masses of the positive- and negative-parity chiral partners are given by
\begin{equation}
    m^d_\pm = \frac{1}{2} \left(\sqrt{a_d^2 \sigma^2 + 4\left(m_0^d\right)^2} \mp b_d\sigma \right) \rm,
\end{equation}
where $\pm$ sign denotes parity and $d \in \lbrace N,\Delta \rbrace$ labels the nucleonic and $\Delta$ doublets. The states $N_\pm$ are identified as $N(939)$ and $N(1535)$, and $\Delta_\pm$ as $\Delta(1232)$ and $\Delta(1700)$~\cite{ParticleDataGroup:2022pth}. We note that other possibilities for identifying the chiral partners are also possible (see, e.g., Refs.~\cite{Zschiesche:2006zj, Jido:1998av}. The parameters $a_d$ and $b_d$ are combinations of Yukawa couplings with the $\sigma$ meson. For a given doublet, they are determined by the corresponding vacuum masses $m_\pm^d$ (see Table~\ref{tab:input_table}).

\begin{table}[t!]
    \centering
    \begin{tabular}{|c|c|c|} \hline
        $n_0~[\rm fm^{-3}]$ & $E/A-m_{N_+}~\rm[GeV]$ & $K~\rm[GeV]$ \\ \hline\hline
        0.16                & $-0.016$                 & 0.24 \\ \hline 
    \end{tabular}
    \caption{Properties of the nuclear ground state at $T=0$ and $\mu_B=0.923~$GeV used in this work: saturation density $n_0$, binding energy $E/A-m_{N_+}$, and incompressibility $K$.}
    \label{tab:gs}
\end{table}

\begin{table}[t!]
    \centering
    \begin{tabular}{|c|c|c|c|c|c|c|c|c|}\hline
         $m_0^N$ & $m_{N^+}$ & $m_{N^-}$ & $m_0^\Delta$ & $m_{\Delta^+}$ & $m_{\Delta^-}$ & $m_\pi$ & $m_\omega$ & $f_\pi$ \\\hline\hline
         0.8 & 0.939     &  1.5    & 0.95 & 1.232          & 1.7          & 0.14   & 0.783      & 0.093   \\ \hline
    \end{tabular}
    \caption{Physical vacuum inputs and the parity doublet model parameters used in this work. All entries are in GeV.}
    \label{tab:input_table}
\end{table}

The parameters of the $\Delta$ sector are generally poorly constrained~\cite {Maslov:2015msa}. The most commonly used constraint is the depth of the $\Delta$ optical potential, $U_\Delta$, at nuclear saturation density. However, there is no clear consensus in the literature as to what its value should be. For example, analyses of the threshold for $\pi$ and $\Delta$ production in heavy ion collisions give $U_N \lesssim U_\Delta \lesssim 2/3 U_N$ (see, e.g.,~\cite{Song:2015hua, Guo:2015tra,Cozma:2014yna,Ferini:2005del}). However, a recent comparison of the optical potentials of the nucleon and $\Delta$ in electron scattering on nuclear targets suggests that $U_\Delta \simeq 3/2 U_N$~\cite{Bodek:2020wbk}. In principle, the values of the ratio $r_\Delta$ and the chirally invariant mass $m_0^\Delta$ in the parity doublet model can be constrained by analyzing the electromagnetic excitations of $\Delta$. This was done in the context of the relativistic quantum hadrodynamics scheme, which constrains the baryon-meson couplings~\cite{Wehrberger:1989cd}. A similar analysis could be performed in the parity doublet model. This would yield a relation between $r_\Delta$ and $m_0^\Delta$. However, such a study is beyond the scope of this paper, and we plan to explore these issues elsewhere. 

\begin{figure}[t!]
    \centering
    \includegraphics[width=1.0\linewidth]{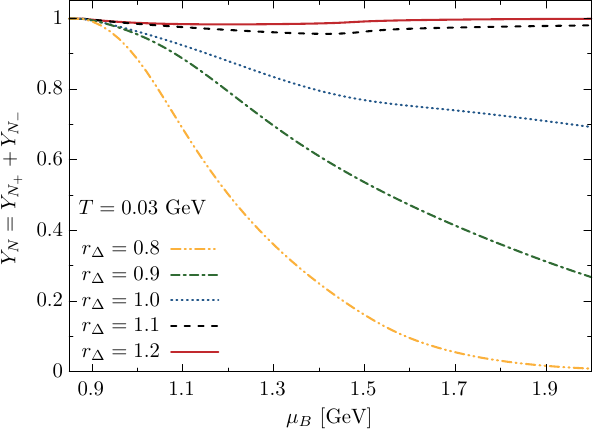}
    \caption{Nuclear matter fraction $Y_N = Y_{N_+} + Y_{N_-}$ as a function of baryon chemical potential at $T=0.03~\rm GeV$ for different values of the repulsive coupling $r_\Delta$. We note that the corresponding $
    \Delta$ matter fraction $Y_\Delta = Y_{\Delta_+} + Y_{\Delta_-} = 1 - Y_N$.}
    \label{fig:R_dep_frac}
\end{figure}

In this work, we focus on a qualitative description of the fluctuations of the net-baryon number density. We consider a representative value of $m_0^N= 0.8~\rm GeV$. This large value is motivated by the recent lattice QCD results~\cite{Aarts:2017rrl, Aarts:2018glk, Aarts:2015mma}. Similar values have also been used in the literature~\cite{Dexheimer:2007tn,Gallas:2009qp,Paeng:2011hy, Sasaki:2011ff, Gallas:2011qp, Zschiesche:2006zj, Benic:2015pia, Marczenko:2017huu, Marczenko:2018jui, Marczenko:2019trv, Marczenko:2020wlc, Marczenko:2020jma, Marczenko:2023ohi, Mukherjee:2017jzi, Mukherjee:2016nhb, Dexheimer:2012eu, Steinheimer:2011ea, Weyrich:2015hha, Sasaki:2010bp, Yamazaki:2018stk,Yamazaki:2019tuo, Ishikawa:2018yey, Steinheimer:2010ib, Giacosa:2011qd,Motohiro:2015taa,Minamikawa:2020jfj,Gao:2024chh}. Note, however, that the qualitative structure of the results presented in this paper does not depend on the choice of $m_0^N$. There are two remaining parameters, $m_0^\Delta$ and $r_\Delta$. At vanishing baryon chemical potential, the EOS does not depend on $r_\Delta$ due to the vanishing $\omega$ at $\mu_B=0$\footnote{We note that while the EOS at vanishing baryon chemical potential and finite temperature, i.e., $p(\eps)$ is not affected by the values of repulsive couplings, the fluctuations depend on them. See, for example,~\cite{Koch:2023oez}.}. Therefore, we first use the finite-temperature EOS at $\mu_B=0$ to determine $m_0^\Delta$. We find that for the chosen value of $m_0^N$, the model yields a smooth chiral crossover transition for $m_0^\Delta \gtrsim 0.95~\rm GeV$. We also require $m_0^N \lesssim m_0^\Delta$. In general, too low values of $m_0^\Delta$ lead to a first-order phase transition at $\mu_B=0$, and also lead to the appearance of $\Delta$ matter at subsaturation densities; thus, spoiling the ground state properties~\cite{Takeda:2017mrm}. On the other hand, setting $m_0^\Delta \rightarrow \infty$ suppresses the $\Delta$ states, and the EOS effectively corresponds to the pure nucleonic EOS. In the present work, we choose $m_0^\Delta = 0.95~\rm GeV$ because it yields the lowest temperature of the chiral crossover transition at the vanishing baryon chemical potential. In Fig.~\ref{fig:sigma_T}, we plot the normalized expectation value of the $\sigma$ mean field at vanishing baryon chemical potential for the pure nucleonic (no $\Delta$) and $m_0^\Delta = 0.95~$GeV models. Including $\Delta$ lowers the pseudocritical temperature from $T_c \simeq 0.21~$GeV to $T_c \simeq 0.17~$GeV. With $m_0^\Delta$ fixed, we are left with a single parameter, $r_\Delta$, and we systematically study the influence of repulsion on the fluctuations of the net-baryon number density. The phase structure at vanishing baryon chemical potential is independent of the strength of the repulsive couplings. Therefore, the strongest dependence of the EOS on $r_\Delta$ is expected at low temperature and high density. 

In Fig.~\ref{fig:phase_diagram}, we show the phase diagram of the parity doublet model for $r_\Delta =0.8,~1.2$. The liquid-gas crossover line obtained at a minimum of $\pp \sigma /\pp \mu_{N_+}$ (see Sec.~\ref{sec:fluct} for details) is almost independent of the choice of $r_\Delta$. This is to be expected since the parameters are chosen so as not to affect the properties of the nuclear ground state at low temperatures and $r_\Delta$ does not affect the EOS at vanishing baryon chemical potential. Other lines, obtained from the corresponding inflection points of $\pp \sigma / \pp \mu_\alpha$, are sensitive to the value of $r_\Delta$ at low temperatures. They converge and continue almost as a single line with increasing temperature until $T_c$ is reached at vanishing baryon chemical potential. In this work, we pay particular attention to the low temperature and high baryon chemical potential part of the phase diagram. We systematically study the influence of $\Delta$ matter on the fluctuations of the net-baryon number in dense matter. We remark that the liquid-gas crossover transition develops a critical point at $T\simeq 0.016~$GeV and turns into an ordinary first-order phase transition. In principle, the existence of critical points at the other crossover lines shown in Fig.~\ref{fig:phase_diagram} depends on the model parametrization. In this case, additional effects, such as non-equilibrium spinodal decomposition would have to be addressed. These effects have been explored, for example, in the context of Nambu--Jona-lasinio model~\cite{Sasaki:2007db, Sasaki:2007qh}. However, this is beyond the scope of the current work. We plan to elaborate on this interesting issue elsewhere.\\

\begin{figure}[!t]
    \centering
    \includegraphics[width=\linewidth]{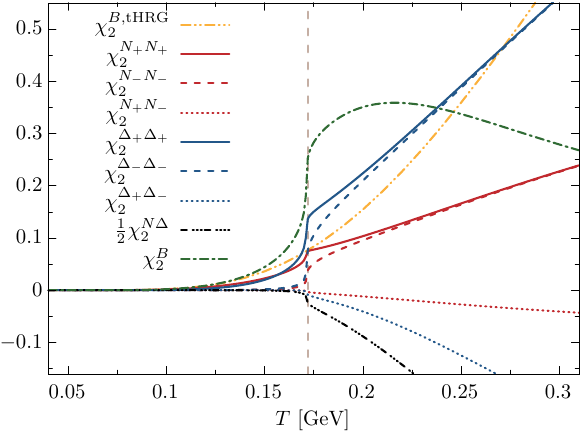}
    \caption{Susceptibilities, $\chi_2^{\alpha\beta}$, as functions of temperature at vanishing baryon chemical potential. Also shown is the net-baryon number susceptibility $\chi_2^{B}$ and the corresponding result, $\chi_2^{B, \rm tHRG}$ obtained in the tHRG model. Note that the correlations between $N_\pm$ and $\Delta_\pm$ are combined in $1/2\chi_2^{N\Delta}$ (see text for details). The vertical, dotted line marks the chiral phase transition.}
    \label{fig:eos_T}
\end{figure}

We define the particle fraction as follows
\begin{equation}\label{eq:fraction}
    Y_\alpha = \frac{n_\alpha}{n_B} \rm,
\end{equation}
where
\begin{equation}\label{eq:density_def1}
    n_\alpha = - \frac{\partial \Omega_\alpha}{\partial \mu_B} \;\;\;\; \textrm{and} \;\;\;\;  n_B = \sum_\alpha n_\alpha \rm,
\end{equation}
where $\Omega_\alpha$ is the kinetic part of the thermodynamic potential for particle species $\alpha$, and $n_\alpha$, $n_B$ are the net density of particle species $\alpha$ and the total net-baryon number density, respectively. Eq.~\eqref{eq:fraction} yields the following constraint $\sum_\alpha Y_\alpha = 1$. In Fig.~\ref{fig:R_dep_frac}, we show the fraction of nuclear matter $Y_{N} \equiv Y_{N_+} + Y_{N_-}$ at $T=0.03~$ GeV for different values of $r_\Delta$. Note that the corresponding fraction of $\Delta$ matter is $Y_\Delta = 1 - Y_N$. As indicated in~\cite{Marczenko:2022hyt}, the asymptotic high-density matter composition depends on the difference of the chemical potentials:
\begin{equation}
\begin{split}
    \mu_N - \mu_\Delta &= g_\omega^N \omega \left(r_\Delta - 1 \right) \\
    &= \left(\frac{g^N_\omega}{m_\omega}\right)^2\left(n_N + r_\Delta n_\Delta\right)\left(r_\Delta-1\right)\rm,
    \end{split}
\end{equation}
where $n_N = n_{N_+} + n_{N_-}$, $n_\Delta = n_{\Delta_+} + n_{\Delta_-}$. Note that the last equality is valid because of stationary conditions. The sign of the right-hand side depends only on the sign of $\left(r_\Delta-1\right)$. Therefore, the composition of asymptotic matter depends solely on the value of $r_\Delta$. For $r_\Delta = 1$, at large $\mu_B$, the composition of the system can be determined by degeneracy factors, that is, $Y_\alpha = \gamma_\alpha / \sum_i \gamma_i$. In that case, $Y_N \rightarrow 0.2$ at high $\mu_B$. For $r_\Delta > 1$, the system is dominated by nuclear matter, that is, $Y_N \rightarrow 1$ at high $\mu_B$. Interestingly, small deviations of $r_\Delta$ from unity produce different compositions of matter at $\mu_B \lesssim 2~$ GeV. For example, for $r_\Delta = 1.1$, we observe that the system is dominated to a large extent by nuclear matter with $Y_N \gtrsim 0.95$. Thus, for $r_\Delta \gtrsim 1.1$, the EOS at low temperatures should already reflect the pure nucleonic EOS. Generally, for $r_\Delta > 1$ the fraction of nuclear matter features a minimum, which is dictated by asymptotic behavior. On the other hand, when $r_\Delta < 1$, the system becomes dominated by $\Delta$ matter and $Y_N \rightarrow 0$. The lower the value of $r_\Delta$, the faster the nuclear matter is suppressed. We find that for $r_\Delta = 0.8 - 1.2$, the fraction of nuclear matter can range from almost zero to unity for $\mu_B \lesssim 2.0~\rm GeV$. We note that $r_\Delta \rightarrow \infty$ and/or $m_0^\Delta \rightarrow \infty$ suppresses $\Delta$ matter and the EOS effectively corresponds to the pure nucleonic EOS.

In this work, we fix $m_0^N = 0.8~\rm GeV$, $m_0^\Delta = 0.95~\rm GeV$, and consider three representative values of $r_\Delta = 0.8,~1.0,~1.2$. We systematically study the influence of $\Delta$ matter on the susceptibility of the net-baryon number density.

\begin{figure}[!t]
    \centering
    \includegraphics[width=\linewidth]{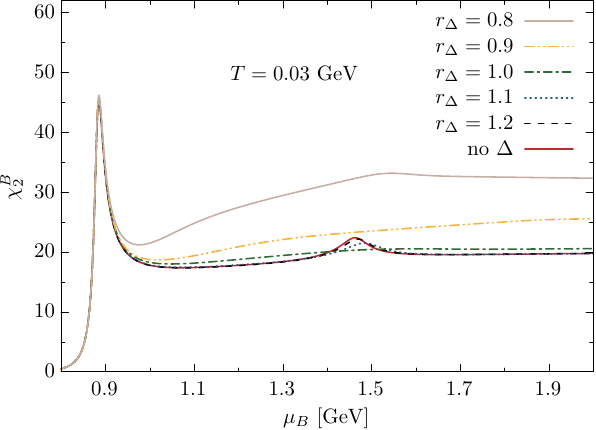}
    \caption{Net-baryon number susceptibility for different values of $r_\Delta$ at $T=0.03~$GeV as functions of the baryon chemical potential.}
    \label{fig:x2}
\end{figure}

\section{Susceptibilities of the net-baryon number density}\label{sec:fluct}

\begin{figure*}
    \centering
    \includegraphics[width=1.0\linewidth]{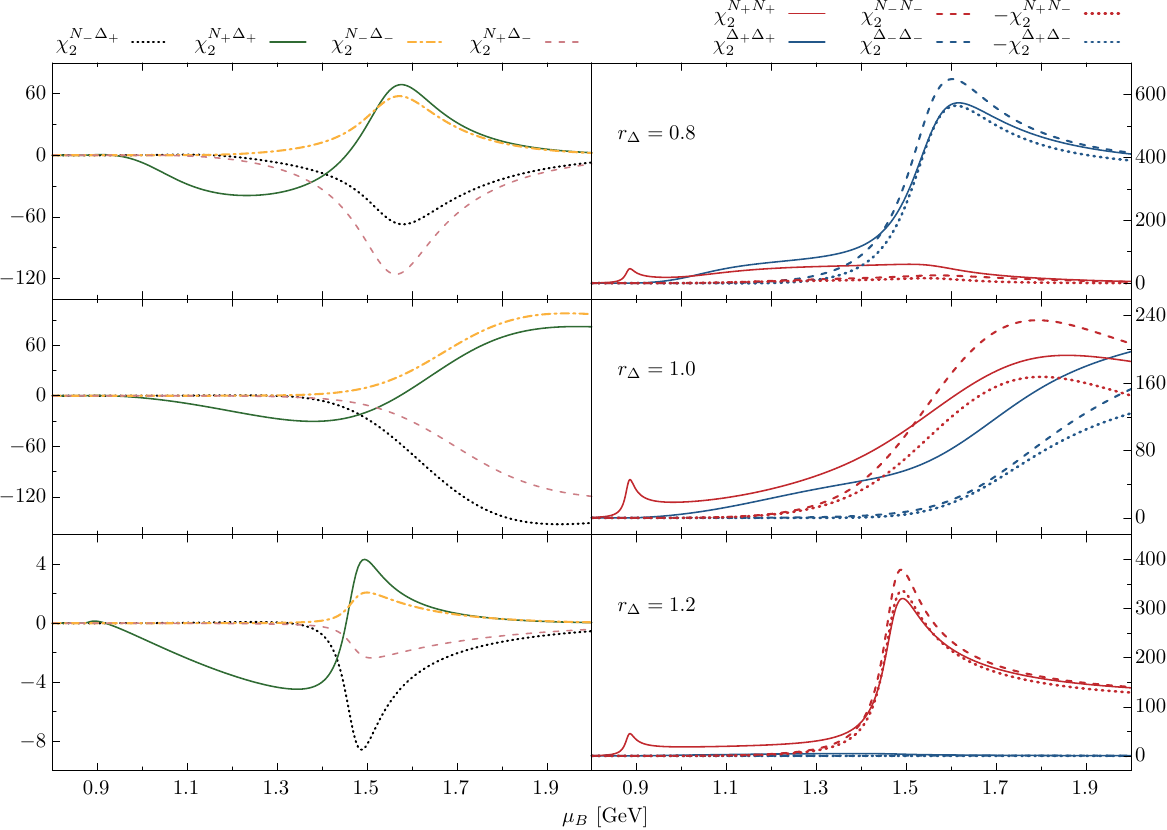}
    \caption{Susceptibilities $\chi_{\alpha\beta}$ for different $r_\Delta$ at $T=0.03~$GeV. Terms within the nucleonic and $\Delta$ sectors are shown in the right panel. The correlations between $N_\pm$ and $\Delta_\pm$ are shown in the left panel. Note that in the right panel, the correlators are shown with a negative sign and that the scale are different in each panel.}
    \label{fig:x2_all}
\end{figure*}

In this work, we assume a system composed of positive and negative parity nucleons and $\Delta$'s. Since isospin correlations are expected to be small~\cite{Fukushima:2014lfa}, for simplicity, we assume isospin symmetry. Thus, each isospin state contributes equally, which is expressed in the degeneracy factors (see Sec.~\ref{sec:pd_model}). For such a system, the net-baryon number is
\begin{equation}
    N_B = N_{N_+} + N_{N_-} + N_{\Delta_+} + N_{\Delta_-}\rm,
\end{equation}
with the $\pm$ sign denoting the net number of positive/negative-parity baryons. The mean and variance can be expressed in terms of cumulants as
\begin{equation}
\begin{split}
   \langle N_B \rangle \equiv \kappa_1^B &= \sum_{\alpha} \kappa_1^\alpha\rm,\\
   \langle \delta N_B \delta N_B\rangle \equiv \kappa_2^B &= \sum_{\alpha,\beta} \kappa_2^{\alpha\beta}\rm,
\end{split}
\end{equation}
respectively. The individual terms in the equations above are
\begin{equation}
\begin{split}
    \kappa_1^\alpha &= \langle N_\alpha \rangle \rm, \\
    \kappa_2^{\alpha\beta} &= \langle\delta N_\alpha \delta N_\beta\rangle \rm,
\end{split}
\end{equation}
where $\alpha, \beta \in \lbrace N_\pm, \Delta_\pm\rbrace$.

In general, the cumulants of the baryon number are defined as
\begin{equation}
    \kappa_n^B \equiv T^n \frac{\dd^n \log{\mathcal Z}}{\dd \mu_B^n}\Bigg|_{T=\it const} \rm,
\end{equation}
where $\mathcal Z$ is the partition function. Because the thermodynamic potential $\Omega$ is related to the grand-canonical partition function through $\Omega = -T\log{\mathcal Z} / V$, one may relate the cumulants with the susceptibilities of the net-baryon number in the following way
\begin{equation}\label{eq:def_x2_kn}
    \kappa_n^B = V T^3\chi_n^B \rm,
\end{equation}
where $V$ is the volume of the system and 
\begin{equation}\label{eq:chi_def}
    \chi_n^B \equiv -\frac{\dd^n \hat\Omega}{\dd \hat\mu_B^n}\Bigg|_T \rm,
\end{equation}
with $\hat \Omega = \Omega/T^4$ and $\hat \mu_B = \mu_B/T$.

\begin{figure}[t!]
    \centering
    \includegraphics[width=\linewidth]{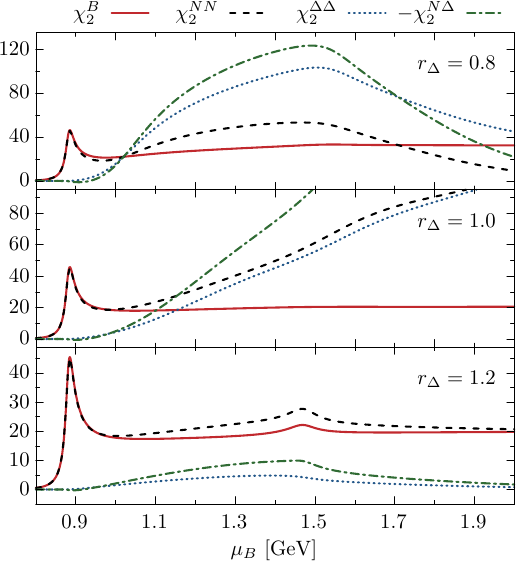}
    \caption{Total susceptibilities within nucleonic and $\Delta$ sectors and correlation between them as defined in Eqs.~\eqref{eq:x2_sectors_def} as functions of the baryon chemical potential at $T=0.03~\rm GeV$. Note that $\chi_{2}^{N\Delta}$ is plotted with negative sign. The total net-baryon number susceptibility is shown with red lines.}
    \label{fig:x2_sectors}
\end{figure}

To be able to connect the individual cumulants $\kappa_n^{\alpha\beta}$ to susceptibilities (as in Eq.~\eqref{eq:def_x2_kn}), we need to rewrite the mean-field thermodynamic potential in terms of the newly defined chemical potentials, $\mu_{N_\pm}$ and $\mu_{\Delta_\pm}$ for positive- and negative-parity nucleons and $\Delta$'s:
\begin{equation}\label{eq:thermo_pm}
\begin{split}
    \Omega &=\Omega_{N_+}\left(\mu_{N_+}, T, \sigma, \omega\right) + \Omega_{N_-}\left(\mu_{N_-}, T, \sigma, \omega\right)\\
    &+\Omega_{\Delta_+}\left(\mu_{\Delta_+}, T, \sigma, \omega\right) + \Omega_{\Delta_-}\left(\mu_{\Delta_-}, T, \sigma, \omega\right)\\
    &+ V_\sigma(\sigma) + V_\omega(\omega)\rm,
\end{split}
\end{equation}
where the mean fields $\sigma = \sigma\left(\mu_{N_+}, \mu_{N_-}, \mu_{\Delta_+}, \mu_{\Delta_-}, T\right)$ and $\omega = \omega\left(\mu_{N_+}, \mu_{N_-}, \mu_{\Delta_+}, \mu_{\Delta_-}, T\right)$. Such a separation into separate chemical potentials is possible in the mean field approximation which is a single particle theory (see detailed discussion in~\cite{Garcia:2018iib}). To be thermodynamically consistent, we need to set each $\mu_\alpha = \mu_B - g^\alpha_\omega \omega$ at the end of the calculations and before the numerical evaluation. We note that $\mu_\alpha$'s are independent variables. The net-baryon density is then given as
\begin{equation}
    n_B = n_{N_+} + n_{N_-} + n_{\Delta_+} + n_{\Delta_-}\rm,
\end{equation}
where $n_\alpha$ are the net densities given by
\begin{equation}\label{eq:dens_pm}
n_\alpha = - \frac{\pp \Omega}{\pp \mu_\alpha}\rm,
\end{equation}
After a little bit of algebra, it can be seen that this definition is consistent with Eq.~\eqref{eq:density_def1}. We stress that the derivative should be taken at a constant temperature.

\begin{figure*}[t!]
    \centering
    \includegraphics[width=\linewidth]{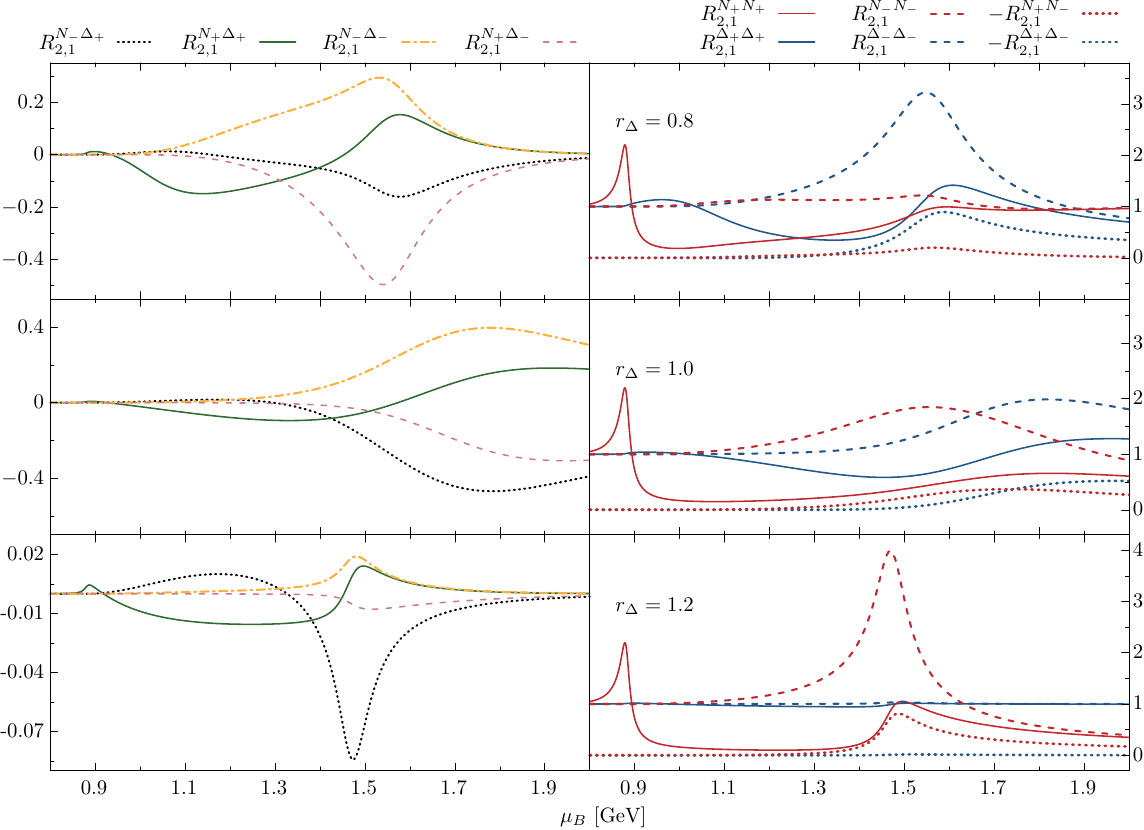}
    \caption{Scaled variances, $R_{2,1}^{\alpha\beta}$, for different sectors and different values of $r_\Delta$ at $T=0.03~$GeV. Ratios within the nucleonic and $\Delta$ sectors are shown in the right panel. The correlations between $N_\pm$ and $\Delta_\pm$ are shown in the left panel. Note that in the right panel the correlators, $R_{2,1}^{N_+N_-}$ and $R_{2,1}^{\Delta_+\Delta_-}$, are shown with negative sign.}
    \label{fig:R21_all}
\end{figure*}

The second-order susceptibility can be expressed as follows
\begin{equation}\label{eq:x2_sum}
\chi_2^B = \chi_2^{NN} + \chi_2^{\Delta\Delta} + \chi_2^{N\Delta}\rm,
\end{equation}
where
\begin{equation}\label{eq:x2_sectors_def}
\begin{split}
    \chi_2^{NN} &= \chi_2^{N_+N_+} + \chi_2^{N_-N_-} +  2\chi_2^{N_+N_-}\rm, \\
    \chi_2^{\Delta\Delta} &= \chi_2^{\Delta_+\Delta_+} + \chi_2^{\Delta_-\Delta_-} + 2\chi_2^{\Delta_+\Delta_-} \rm, \\
    \chi_2^{N\Delta} &= 2\left[\chi_2^{N_+\Delta_+} + \chi_2^{N_+\Delta_-} + \chi_2^{N_-\Delta_+} + \chi_2^{N_-\Delta_-}\right]\rm,
\end{split}
\end{equation}
which contain contributions from the nucleonic, $\Delta$ sectors, and terms that mix them. We note that in~\cite{Koch:2023oez}, a simplified system consisting of a single parity doublet was considered. By explicitly including the $\Delta_\pm$ states, in addition to the individual nucleonic and $\Delta$ sectors, four additional susceptibilities that mix them have to be considered. The individual terms in Eq.~\eqref{eq:x2_sectors_def} equation are given as follows
\begin{equation}\label{eq:chi_2_ab}
    \chi_2^{\alpha\beta} =  \frac{1}{VT^3} \kappa_2^{\alpha\beta} = -\frac{\dd^2 \hat\Omega}{\dd \hat\mu_\alpha \dd\hat\mu_\beta}\Bigg|_{T=\it const} \rm,
\end{equation}
where $\hat\mu_{x} = \mu_{x}/T$. The detailed derivation of the cumulants $\kappa_2^{\alpha\beta}$ is presented in~\cite{Koch:2023oez} for the system consisting of nucleonic parity partners. The inclusion of $\Delta_\pm$ is straightforward. We notice that, in the mean field approximation, $\chi_2^{\alpha\beta} = \chi_2^{\beta\alpha}$. 

The susceptibilities defined in Eq.~\eqref{eq:x2_sectors_def} can be related through Eq.~\eqref{eq:chi_2_ab} to the cumulants of the two-component system:
\begin{equation}\label{eq:cumulants_sectors}
\begin{split}
    \kappa_2^{NN} &= \langle \delta N \delta N\rangle \rm,\\
    \kappa_2^{\Delta\Delta} &= \langle \delta \Delta \delta \Delta\rangle \rm,\\
    \kappa_2^{N\Delta} &= \langle \delta N \delta \Delta\rangle \rm,\\
\end{split}
\end{equation}
with
\begin{equation}
    \kappa_2^B = \kappa_2^{NN} + \kappa_2^{\Delta\Delta} + 2 \kappa_2^{N\Delta} \rm,
\end{equation}
where the nucleon, $\Delta$ and net baryon numbers are  $N = N_+ + N_-$, $\Delta = \Delta_+ + \Delta_-$, and $N_B = N + \Delta$, respectively.

Event-by-event cumulants and correlations are extensive quantities. They depend on the volume of the system and its fluctuations, which are unknown in heavy-ion collisions. However, the volume dependence can be canceled by taking the ratio of cumulants. Therefore, it is useful to define the scaled variance, i.e., the ratio of the second- to first-order cumulants of the baryon number, which may also be expressed through susceptibilities,
\begin{equation}
    R_{2,1}^B \equiv \frac{\kappa_2^B}{\kappa_1^B} = \frac{\chi_2^B}{\chi_1^B} \rm.
\end{equation}
To analyze the scaled variances for individual particle species we also define the partial scaled variances:
\begin{equation}\label{eq:r21_def}
    R_{2,1}^{\alpha\beta} \equiv \frac{\kappa_2^{\alpha\beta}}{\sqrt{\kappa_1^\alpha \kappa_1^{\beta}}} = \frac{\chi_2^{\alpha\beta}}{\sqrt{\chi_1^\alpha \chi_1^\beta}} \rm.
\end{equation}
The scaled variances, $R_{2,1}^{\alpha\beta}$, can be defined also for cumulants related to the nucleonic and $\Delta$ sectors in Eq.~\eqref{eq:cumulants_sectors}. We note that, in general, the partial scaled variances $R_{2,1}^{\alpha\beta}$, are not additive. For example, $R_{2,1}^{NN} + R_{2,1}^{\Delta\Delta} + 2R_{2,1}^{N\Delta} \neq R_{2,1}^B$. We note that due to experimental limitations, the net-proton number fluctuations are assumed to reflect the fluctuations of the net-baryon number. However, this relation has not yet been explored in theoretical models with dynamical chiral symmetry restoration.

The susceptibilities introduced in Eq.~\eqref{eq:chi_2_ab}, can be evaluated analytically by differentiating Eq.~\eqref{eq:thermo_pm}. Explicit calculations yield
\begin{equation}\label{eq:x2_ab2}
\begin{split}
    \chi_2^{\alpha\beta} = &-\frac{\pp \sigma}{\pp \hat\mu_\beta} \left( \frac{\pp^2 \hat\Omega}{\pp \sigma^2}\frac{\pp \sigma}{\pp \hat\mu_\alpha} + \frac{\pp^2\hat\Omega}{\pp\sigma\pp\omega}\frac{\pp\omega}{\pp\hat\mu_\alpha} - \frac{\pp \hat n_\alpha}{\pp\sigma} \right)\\
    &-\frac{\pp\omega}{\pp\hat\mu_\beta} \left( \frac{\pp^2\hat\Omega}{\pp\omega^2}\frac{\pp \omega}{\pp\hat\mu_\alpha} + \frac{\pp^2\Omega}{\pp\sigma\pp\omega}\frac{\pp\sigma}{\pp\hat\mu_\alpha} - \frac{\pp \hat n_\alpha}{\pp\omega}\right)\\
    &+\frac{\pp\sigma}{\pp\hat \mu_\alpha}\frac{\pp \hat n_\beta}{\pp\sigma}+ \frac{\pp\omega}{\pp\hat \mu_\alpha}\frac{\pp\hat n_\beta}{\pp\omega} + \frac{\pp \hat n_\alpha}{\pp\hat \mu_\beta} \rm,
    \end{split}
\end{equation}
where $\hat n_{\alpha/\beta} = n_{\alpha/\beta} /T^3$, and $n_{\alpha/\beta}$ are the net densities defined in Eq.~\eqref{eq:dens_pm}. We note that the last term, $\pp \hat n_\alpha / \pp\hat\mu_\beta = 0$ for $\alpha \neq \beta$.

To evaluate Eq.~\eqref{eq:x2_ab2}, one needs to extract the derivatives of the mean fields w.r.t effective chemical potentials. The derivatives can be calculated out by differentiating the gap equations, namely
\begin{equation}
    \frac{\dd}{\dd \hat\mu_\alpha}\left(\frac{\pp \hat\Omega}{\pp \sigma}\right)\Bigg|_{T=\it const} = \frac{\dd}{\dd \hat\mu_\alpha}\left(\frac{\pp \hat\Omega}{\pp \omega}\right)\Bigg|_{T=\it const} = 0 \rm.
\end{equation}
Writing them explicitly and isolating $\pp \sigma / \pp \hat\mu_\alpha$, $\pp \omega / \pp \hat\mu_\alpha$, yields
\begin{equation}
\begin{split}
    \frac{\pp \sigma}{\pp \hat \mu_\alpha} = &\left( \frac{\frac{\pp^2 \hat \Omega}{\pp\sigma\pp\omega}}{\frac{\pp^2\hat\Omega}{\pp\omega^2}}\frac{\pp\hat n_\alpha}{\pp\omega} - \frac{\pp \hat n_\alpha}{\pp\sigma} \right) \Bigg/ \left(\frac{\pp^2\hat\Omega}{\pp\sigma^2} - \frac{\left(\frac{\pp^2\hat\Omega}{\pp\sigma\pp\omega}\right)^2}{\frac{\pp^2\hat\Omega}{\pp\omega^2}}\right)\rm,\\
    \frac{\pp\omega}{\pp\hat\mu_\alpha} = &-\left(\frac{\pp \hat n_\alpha}{\pp\omega} + \frac{\pp^2\hat\Omega}{\pp\sigma\pp\omega}\frac{\pp\sigma}{\pp\hat \mu_\alpha}\right) \Bigg/ \frac{\pp^2\hat\Omega}{\pp\omega^2} \rm.
\end{split}
\end{equation}
We note that corresponding derivatives of the mean fields w.r.t. $\hat\mu_\beta$ can be found similarly upon replacing $\alpha \rightarrow \beta$. The above derivatives can be plugged into Eq.~\eqref{eq:x2_ab2}. Once the values of the mean fields are established, Eq.~\eqref{eq:x2_ab2} can be evaluated numerically.

\section{Results}\label{sec:results}

First, we study the susceptibilities at finite temperatures and vanishing baryon chemical potential. They are shown in Fig.~\ref{fig:eos_T}, for $r_\Delta=0.8$. It is instructive to compare our results with those of the truncated hadron resonance gas (tHRG) model, where the only degrees of freedom are nucleons, $\Delta$s, and their chiral partners. The HRG model is widely used for the description of matter under extreme conditions, e.g., in the context of heavy-ion collision phenomenology~\cite{Karsch:2003vd, Karsch:2003zq, Karsch:2013naa, Andronic:2012ut, Albright:2014gva, Albright:2015uua}. The thermodynamic potential of the tHRG model is a mixture of uncorrelated ideal gases of stable particles:
\begin{equation}\label{eq:hrg_thermo}
    \Omega^{\rm tHRG} = \sum_{\alpha} \Omega_\alpha \textrm,
\end{equation}
where $\alpha = \lbrace N_\pm, \Delta_\pm \rbrace$ with $\Omega_\alpha$ given by Eq.~\eqref{eq:kinetic_thermo}. The masses are taken to be the vacuum masses (see Table~\ref{tab:input_table}) and $\mu_N = \mu_\Delta = \mu_B$. The net-baryon density and its susceptibility are obtained through Eq.~\eqref{eq:chi_def}. Thus, in the tHRG model one has,
\begin{equation}
\chi_2^{B, \rm tHRG} = \chi_2^{N_+N_+} + \chi_2^{N_-N_-} + \chi_2^{\Delta_+\Delta_+} + \chi_2^{\Delta_-\Delta_-}\rm.    
\end{equation}
Due to the uncorrelated nature of $N_\pm$ and $\Delta_\pm$ in the tHRG model, there are no correlation terms in $\chi_2^{B,\rm tHRG}$. The susceptibility in the tHRG model increases monotonically and we do not observe any critical behavior. The result obtained in the parity doublet model deviates from the tHRG baseline due to the in-medium properties of the baryons. It reaches its maximum above $T_c$ and starts to decrease due to decreasing negative correlations. The positive-parity susceptibilities, $\chi_2^{N_+N_+}$ and $\chi_2^{\Delta_+\Delta_+}$, dominate at low temperatures, increase rapidly as $T_c$ is approached and continue to increase above it. The negative-parity susceptibilities, $\chi_2^{N_-N_-}$ and $\chi_2^{\Delta_-\Delta_-}$, become non-negligible only near $T_c$ and converge to their positive-parity counterparts at higher temperatures, due to chiral symmetry restoration. Similarly, the correlators become relevant only in the vicinity of $T_c$. Generally, the correlations are negative at vanishing baryon chemical potential. We note that although the susceptibilities at vanishing baryon chemical potential are not independent of $r_\Delta$, they are qualitatively similar.

We remark on the simplified nature the current model. Namely, the $N(1535)$, $\Delta(1232)$, and $\Delta(1700)$ resonances are considered to be stable particles by neglecting their finite widths. However, including the effects of finite width in a self-consistent way is non-trivial in the context of the relativistic mean-field approach. Forecasting possible effects on fluctuation observables would require to account for the imaginary part of the resonance self-energy as done, e.g., for $N_-(1535)$ in dense nuclear matter~\cite{Suenaga:2017wbb}.

\begin{figure}[t!]
    \centering
    \includegraphics[width=\linewidth]{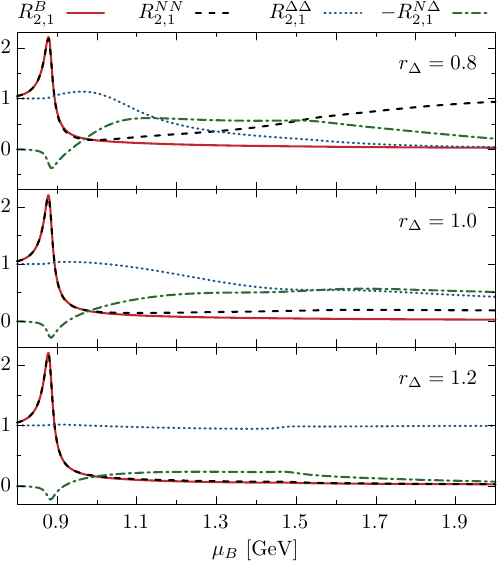}
    \caption{Scaled variances for different sectors and different values of $r_\Delta$ at $T=0.03~$GeV. Note that $R_{2,1}^{N\Delta}$ is plotted with a negative sign. Also shown is the ratio $R_{2,1}^B$ for the net-baryon number susceptibility (solid, red line).}
    \label{fig:R21_sectors}
\end{figure}

Now we turn to the finite baryon chemical potential regime and consider fixed $T=0.03~\rm GeV$. In Fig.~\ref{fig:x2}, we show the net-baryon number susceptibility. The behavior at $\mu_B < 1~\rm GeV$ is similar for all the values of $r_\Delta$ shown in the figure. The curves show a well-pronounced peak which is a remnant of the liquid-gas phase transition. Small deviations from the pure nucleonic result are due to a small fraction of the $\Delta$ matter present at $T=0.03~$GeV. At higher chemical potentials, the susceptibilities develop plateaux. A small peak is seen for the pure nucleonic EOS around $\mu_B=1.4~\rm GeV$, which is a remnant of the first-order chiral phase transition at low temperatures. The secondary peak for $r_\Delta=0.8$ is a result of lower repulsion for the $\Delta$ matter, leading to an early onset of the $\Delta_+$ state. The result for $r_\Delta=1.1$ is very close to the pure nucleonic EOS. We note that for $r_\Delta = 1.2$, the net-baryon number susceptibility converges to that of pure nucleonic EOS. This convergence is due to the suppression of $\Delta$ matter by repulsive interactions. However, the net-baryon number susceptibility at higher baryon chemical potential shows very little sensitivity to the choice of $r_\Delta$ and depends only on the composition of the asymptotic matter, as discussed in the previous section.

The individual susceptibilities $\chi_2^{\alpha\beta}$ also depend on the values of repulsive couplings. This can be seen in the right panel of Fig.~\ref{fig:x2_all}. For $r_\Delta < 1$, the susceptibility is dominated by the contribution of $\Delta_\pm$. Similarly, for $r_\Delta > 1$, they are dominated by the contribution of $N_\pm$. This is expected from the relative abundances shown in Fig.~\ref{fig:R_dep_frac} and discussed above. For $r_\Delta = 1$ all susceptibilities are not negligible. In general, the susceptibilities $\chi_2^{\alpha\alpha}$ at high baryon chemical potential are much larger than the corresponding peak from the liquid-gas transition below $\mu_B = 1~$GeV. At the same time, the positive-negative parity correlations within the same doublet become negative and show minima of similar size as the peaks in $\chi_2^{\alpha\alpha}$. The correlations between $N_\pm$ and $\Delta_\pm$ are shown in the left panel of Fig.~\ref{fig:x2_all} and have a qualitatively similar structure for different values of $r_\Delta$. However, their magnitude depends on the choice of the repulsive coupling and they are non-negligible, except for $r_\Delta = 1.2$, for which the $\Delta$ matter is almost completely suppressed. Interestingly, we find that the susceptibilities of the opposite parity states (which are not chiral partners), i.e., $\chi_2^{N_+\Delta_-}$ and $\chi_2^{N_-\Delta_+}$ become large and negative. They become dominant and have minima around the chiral crossover, providing an additional source of cancellation of the net-baryon number fluctuations at finite density.

The collective susceptibilities defined in Eq.~\eqref{eq:x2_sectors_def} are shown in Fig.~\ref{fig:x2_sectors}. For $r_\Delta = 1.2$, susceptibilities containing $\Delta$ vanish due to a large repulsive coupling for $\Delta_\pm$. Similarly, for $r_\Delta=0.8$, the system is initially driven by the nucleon susceptibility $\chi_2^{NN}$, but $\chi_2^{\Delta\Delta}$ quickly becomes dominant and $\chi_2^{NN}$ vanishes at a higher baryon chemical potential. For $r_\Delta=1.0$, all terms continue to grow with the baryon chemical potential. On the other hand, for $r_\Delta = 1.2$, the susceptibilities in the $\Delta$ channel and the $N\Delta$ correlations are small due to the suppression of the $\Delta$ matter. We note that the  $\chi_2^{NN}$ and $\chi_2^{\Delta\Delta}$ susceptibilities are mostly positive. On the other hand, as soon as the contribution from $\Delta_\pm$ becomes non-negligible, $\chi_2^{N\Delta}$ becomes significant and negative. The susceptibilities $\chi_2^{NN}$ and $\chi_2^{\Delta\Delta}$ are smaller than the individual susceptibilities shown in the right panel of Fig.~\ref{fig:x2_all}, due to the repulsive nature of the correlations of the opposite-parity chiral partners. The correlations between $N_\pm$ and $\Delta_\pm$ are non-negligible and generally tend to be negative at high baryon chemical potential. Thus, the correlator $\chi_2^{N\Delta}$ is negative at large baryon chemical potentials, further suppressing the susceptibility of the total net-baryon number density.

As shown in Fig.~\ref{fig:R21_all}, qualitatively similar structures can be seen in the scaled variances defined in Eqs.~\eqref{eq:r21_def}. At low chemical potential, the sensitivity to the liquid-gas transition is seen in $R_{2,1}^{N_+N_+}$. At high baryon chemical potential as the chiral symmetry becomes restored, the signal is dominated by the scaled variances of the negative parity states, which develop peaks related to the onset of the chiral partners, $N_-$ and $\Delta_-$. The mixed scaled variances between $N_\pm$ and $\Delta_\pm$ show patterns similar to the susceptibilities $\chi_2^{\alpha\beta}$. At high baryon chemical potential, the ratios $R_{2,1}^{N_+\Delta_+}$ and $R_{2,1}^{N_-\Delta_-}$ feature peaks related to the restoration of chiral symmetry, while $R_{2,1}^{N_-\Delta_+}$ and $R_{2,1}^{N_+\Delta_-}$ feature minima.

Most of the structure seen in the scaled variances is washed out when the full nucleonic and $\Delta$ sectors are considered as defined in Eqs.~\eqref{eq:x2_sectors_def} and \eqref{eq:r21_def}. This is shown in Fig.~\ref{fig:R21_sectors}. For $r_\Delta=0.8$, $R_{2,1}^{NN} \simeq R_{2,1}^B$ below $\mu_B \simeq 0.9~\rm GeV$ until $\Delta$ matter becomes populated. This is also reflected in the enhancement of $R_{2,1}^{\Delta\Delta}$, which converges to $R_{2,1}^B$ with a higher baryon chemical potential. At the same time $R_{2,1}^{NN}$ goes to unity, due to $r_\Delta < 1$. Similarly for $r_\Delta=1.0$, $R_{2,1}^{NN}$ deviates from $R_{2,1}^B$ as soon as the $\Delta$ matter sets in, i.e., $R_{2,1}^{\Delta\Delta}$ deviates from unity. However, since at high density the composition is determined by the degeneracy factors, none of the ratios converge to $R_{2,1}^B$. For $r_\Delta = 1.2$, $R_{2,1}^{\Delta\Delta} \simeq 1$, due to the suppression of $\Delta_\pm$ states. Consequently, $R_{2,1}^{NN} \simeq R_{2,1}^B$. However, $R_{2,1}^B$ is suppressed compared to other scaled variances, regardless of the choice of $r_\Delta$. 

\section{Summary}\label{sec:conclusions}

We have studied the net-baryon number density fluctuations in dense matter. We have used the parity doublet model for the nucleons and $\Delta$ baryons in the mean-field approximation. The model features baryonic chiral partners of opposite parity. It allowed us to study the interplay of the positive- and negative-parity baryons, as well as various correlations between them, and their role in fluctuations of net-baryon number density. In this work, we analyzed the second-order susceptibilities, focusing on the low-temperature and high-density part of the phase diagram.

We confirmed that the fluctuations near the liquid-gas phase transition are dominated by the fluctuations of the positive-parity nucleon. This is not the case for larger baryon chemical potential (or net-baryon density). The contributions of other species become non-negligible, and the structure of the individual susceptibilities and correlations becomes nontrivial. Generally, the total fluctuations in a given baryonic chiral doublet (i.e., the fluctuations of the positive- and negative-parity states and their correlation) are smaller than the individual fluctuations. This is due to the non-vanishing correlation between the chiral partners. Furthermore, the total net-baryon number fluctuations are further suppressed compared to the fluctuations of the individual parity doublets. The source of this suppression comes from the correlations between nucleonic and $\Delta$ sectors. We observe that the susceptibility of the negative-parity states becomes dominant at high baryon chemical potentials. This is even more evident when the scaled variances are considered, i.e., ratios of the second- to first-order susceptibility. Generally, the susceptibilities and correlators of species with the same parity are positive at large baryon chemical potential. On the other hand, the correlators of opposite-parity species become negative with increasing baryon chemical potential. Our results highlight the importance of the correlations of the baryonic parity partners and different baryons for the net-baryon number fluctuations. These qualitative differences in fluctuations and correlations can be useful in the analysis of the experimental data. Therefore, it is essential to consistently incorporate the chiral in-medium effects to fully interpret the properties of dense matter created in the next generation of large-scale nuclear experiments FAIR/GSI and NICA/Dubna.

We expect more refined structures to be present in higher-order susceptibilities and their ratios. It is also crucial to determine the role of the baryonic chiral partners in the correlations of different conserved charges, e.g., baryon and electric charge. Work in these directions is underway and will be reported elsewhere.

\section*{Acknowledgments}
The author acknowledges helpful suggestions and comments from Pasi Huovinen, and discussions with Krzysztof Redlich and Chihiro Sasaki. This work is supported by the Excellence Initiative - Research University Program of the University of Wroc\l{}aw of the Ministry of Education and Science.

\bibliography{biblio}

\end{document}